\documentclass[conference]{IEEEtran}

\usepackage{fancyhdr}
\fancypagestyle{first}{%
	\fancyhead[CH]{International Conference ASONAM 2015}
}
\fancypagestyle{then}{%
  \fancyhead[LH]{Canu et al. - Fast community structure local uncovering...}
  \fancyhead[RH]{ASONAM 2015}
}

\usepackage{amssymb}
\usepackage{stmaryrd}
\usepackage[usenames,dvipsnames,svgnames,table]{xcolor}

\usepackage{cite}

\ifCLASSINFOpdf
  \usepackage[pdftex]{graphicx}

  \graphicspath{{./images/}}
\else
\fi

\usepackage[cmex10]{amsmath}
\DeclareMathOperator*{\argmax}{arg\,max}
\DeclareMathOperator*{\argmin}{arg\,min}

\usepackage{algorithm}
\usepackage{algorithmic}

\usepackage[caption=false,font=footnotesize]{subfig}
\usepackage{url}

\usepackage[english]{babel}
\begin{document}

%
\title{Fast community structure local uncovering by independent vertex-centred process}


\author{\IEEEauthorblockN{Ma\"el Canu\IEEEauthorrefmark{1}\IEEEauthorrefmark{2},
Marcin Detyniecki\IEEEauthorrefmark{1}\IEEEauthorrefmark{2}\IEEEauthorrefmark{3}\IEEEauthorrefmark{4},
Marie-Jeanne Lesot\IEEEauthorrefmark{1}\IEEEauthorrefmark{2} and
Adrien Revault d'Allonnes\IEEEauthorrefmark{5}}

\IEEEauthorblockA{\IEEEauthorrefmark{1}Sorbonne Universit\'es, UPMC Univ Paris 06, UMR 7606, LIP6, F-75005, Paris,
France\\
E-mail: \textless{}first\_name\textgreater{}.\textless{}last\_name\textgreater{}@lip6.fr}
\IEEEauthorblockA{\IEEEauthorrefmark{2}CNRS, UMR 7606, LIP6, F-75005, Paris, France}
\IEEEauthorblockA{\IEEEauthorrefmark{3}Data Innovation Lab, GIE AXA, F-75008, Paris, France}
\IEEEauthorblockA{\IEEEauthorrefmark{4}Polish Academy of Sciences, IBS PAN, Warsaw, Poland}
\IEEEauthorblockA{\IEEEauthorrefmark{5}Universit\'e Paris 8, EA 4383, LIASD, FR-93526, Saint-Denis, France\\
E-mail: allonnes@ai.univ-paris8.fr}
}
\maketitle

\thispagestyle{first}

\begin{abstract}
  This paper addresses the task of community detection and proposes a
  local approach based on a distributed list building, where each
  vertex broadcasts basic information that only depends on its degree
  and that of its neighbours. A decentralised external process then
  unveils the community structure. The relevance of the proposed
  method is experimentally shown on both artificial and real data.

\end{abstract}


\section{Introduction}
Social networks are usually represented as graphs, where each vertex is a person and each edge between two vertices reflects a particular relationship, e.g. these people know each other or share a common interest \cite{wasserman_social_1994}. The structure of a social network graph often reveals the presence of communities, intuitively defined as subsets of vertices more densely connected within themselves than with others.

When interacting in today's real world in this context, most people carry a
wireless communication device --such as a smartphone but also so-called \textit{smart} objects, such as smart watches or wearables to name a few-- and constitute
the vertices of a volatile ad-hoc delay-tolerant communication network known as opportunistic mobile social network, a class of Mobile Ad-hoc NETwork (MANET), and encompasses a
community structure as well \cite{chaintreau_impact_2007,vastardis_mobile_2013}.
The automatic identification of these communities can prove beneficial
to better route the traffic inside that
network~\cite{lee_mobility_2013}.

Hui et al. \cite{hui_distributed_2007} showed that knowing the community structure of such a network can help improving the routing of packet traffic inside it. If the network is formed through human-carried devices, thus reflecting a probable social organisation, the improvement can be significant \cite{chaintreau_impact_2007,hui_bubble_2011}.

Moreover, if the emergence of a network occurs at specific place, such as a museum or a store, or during a specific event such as an exhibit or a job-dating, the live uncovered community structure can be used to offer an enhanced experience to the visitor/customer by recommending specific content or places, and thereby to increase merchant benefits. Experiments conducted at MIT where each user carries a small electronic communicating badge on their shirt show that these badges are useful to successfully live build affinity models among participants and infer interest between people \cite{laibowitz_sensor_2006,paradiso_identifying_2010}. We believe that adding the community structure to the features used to build the models, bringing a social dimension, could improve the models quality.
However, it is a difficult task and a challenging problem in the field
of graph mining, as detailed in Section~\ref{motiv}. Most existing
methods, at the protocol level implementation, require vertices of the network
to share a certain amount of information implying many exchanges between them, or
calculations requiring many inputs.

This paper proposes a simple distributed method with reduced
propagation, therefore easily deployable in the context mentioned above. We show that, only by having vertices share a list based on the degrees of their neighbours, we are able to unveil a coherent community structure.

This paper is organised as follows: the next section discusses related works, Section~\ref{algo} describes the method we propose and
Section~\ref{expes} presents and discusses some experimental results.

\section{Related Works}
\label{motiv}
The community detection task aims at decomposing a graph into
subgraphs that are more densely connected within themselves than with
others. Several formal definitions of community have been
proposed\cite{radicchi_defining_2004,fortunato_community_2009}, though
none is universally accepted, opening the way to many different
approaches. We propose to distinguish between 2~main categories, namely
global and local methods.

\subsection{Global Approaches}

The global approaches, introduced first, initially
derived from computational graph analysis, like partitioning,
clustering or clique-finding
\cite{kernighan_efficient_1970,bron_algorithm_1973,freeman_set_1977}.

The exploitation of graph topology analysis then led to the
introduction of the modularity measure
\cite{girvan_community_2002,newman_finding_2004}, used in numerous algorithms, such as Clauset
\cite{clauset_finding_2004} or Louvain~\cite{blondel_fast_2008} to
name two examples. Other criteria have been proposed to address the
limits of modularity
\cite{aldecoa_deciphering_2011,fortunato_resolution_2007}. Other
approaches include techniques based on statistical
models \cite{hoff_latent_2002} or informations flows
\cite{rosvall_maps_2008} for example. 

However, these methods require to know the entire graph topology,
imposing to dispose of the whole graph for their processing.

\subsection{Local Approaches}

The second category, made of local approaches, detect communities by
means of a local measure, processing only a subset of the graph,
e.g. a subset of vertices, a subset of edges, or a subgraph.  These local
approaches allow an analysis portion by portion, thereby facilitating the
conception of a distributed implementation.  They also make it easier
to process dynamic graphs, as they can adapt to local modifications of the graph
\cite{cazabet_simulate_2011}. In particular, they apply to the case of
opportunistic networks mentioned in the introduction, such as Pocket
Switched Networks \cite{hui_pocket_2005}: in this case, community
detection is difficult using a global approach because there is no
central unit knowing all the network vertices at the same time.

Local approaches can be defined as variations of global methods
described in the previous subsection: modularity has, for instance,
been adapted for such a local exploitation \cite{clauset_finding_2005}.

Other recent local approaches include \textit{label propagation}
\cite{raghavan_near_2007} and \textit{seed-centric} \cite{kanawati_seed-centric_2014}.
Label propagation basically consists in assigning a label
to each vertex and propagating these labels between neighbours and throughout the network
following given rules. After achieving convergence, vertices labelled with the same label
will be considered as belonging to the same community.
However, propagation to the whole network can lead to flooding effects and can
be a significant downside as opportunistic networks are often
delay-tolerant so the synchronisation can be tough.
Seed-centric consists in identifying a vertex or group of vertices
displaying specific properties that will constitute a basis (the \textit{seed}) on
which the community detection is performed.


The local approach we propose, as detailed in the next section, requires little computation,
no centralised process and limits the information exchange to the close neighbours.

%
%

\section{Proposed algorithm}
\label{algo}
\pagestyle{then}
This section presents the algorithm we propose, first exposing the
underlying concepts, then describing each part separately and
examining its properties.

We consider an undirected and unweighted graph without isolated vertices nor self-loops,
denoted $G=(V,E)$, where $V$ is the set of vertices, $E$ the set of
edges, $n = |V|$ and $m = |E|$, $d_v$ is the degree and $\Gamma(v)$ the set of neighbours of a vertex
$v$. The absence of isolated vertex implies that $\forall v \in V, d_v > 0$.

$\mathcal{C} = \{c_1, ..., c_{|\mathcal{C}|}\}$ is the set of
communities formed after detection, and $C(v)$ refers to the community
a vertex~$v$ belongs to.


\subsection{Overview and Architecture}
The principle on which the proposed approach relies is that, in a
context where a vertex can be aware of its environment, like an
electronic communicating device as mentioned above, minimal
information sharing among vertices is sufficient to serve as a basis
to form communities. This method does not optimise a mathematical
property characterizing communities like modularity, but is well
suited for use in a decentralized, mobile networking environment
requiring minimal computation.

The proposed method makes the vertices perform the essential part of the computation:
metaphorically, each vertex answers the question: ``Which of my neighbours am I the most
related to?''. This step allows to identify locally the vertex or set of vertices
attracting most of their neighbours. Note that this assumption has a meaning close to the preferential
attachment theory \cite{barabasi_emergence_1999}.

Then, the community structure uncovering process brings together the vertices that most want to be together. This process is external to the vertices,
so no vertex has to propagate information to others.
At the end of the uncovering process, each vertex knows its own
community, but not the whole community structure of the network. This
structure may be found by doing a graph traversal, or an exploration
of the graph.  The advantage is that only a part of the community
structure can be retrieved, with a partial graph traversal or local
exploration for example, adjusting the computing resources consumption
to just what is needed. Indeed, the entire structure is often not
required, as for the Internet: no router knows the full topology because
it is too vast and too fast-changing. However, a knowledge of a
subpart of the Internet is sufficient for efficiently routing the
packets.

In short, the desired overall effect is first to locally identify the densest areas in the graph where the vertices of highest degree are found, and then to group together vertices around them. This approach is related to the seed-centric community detection method family, except that we do not seek to identify
a seed area but rather to have the vertices know and organise themselves to reveal the community structure.

The proposed method requires little computation power and memory use: the only prerequisite is that a vertex must be able to communicate with its neighbours to get or send minimal information, for example about the degrees. The calculation is then easily distributed over the vertices themselves, i.e. the connected devices constituting network.


More precisely, the algorithm is composed of 3 steps successively detailed in the next subsections:
\begin{enumerate}
\item \label{ph1} List of candidates compiling: each vertex builds a list including some of its neighbours depending on their degree. Each vertex can run this step independently, at the same time.
\item \label{ph2} Agreement computation and assignment: in this step,
  each vertex $v$ shares its list with its neighbourhood. It calculates the proportion of vertices
  in common between its list and those of its neighbours
  (\textit{agreement}). It selects the neighbour $a_v$ it has the highest
  agreement with.
\item \label{ph3} Community uncovering on $\mathcal{V}$: for each pair
  $(v, a_v)~\in~\mathcal{V}$, the communities of $v$ and $a_v$ are
  merged.
\end{enumerate}

\subsection{List of candidates compiling}
\label{listbuild}
We think that having vertices share minimal information to their close neighbourhood is a simple yet efficient way to identify denser areas in the graph, assimilated with communities.
We propose the minimal information required as an input for a vertex $v$ to be the degree of each of its direct
neighbours, so $v$ disposes of $d_v$ pieces of information for input.

Then, each vertex $v$ keeps only its neighbours of highest degree in a list called~$S_v$.

More precisely, we define an integer $k_v$ such that for each~$v$, $k_v~\in~\llbracket 1~...~d_v\rrbracket$, and an ordering function $\sigma_v$ such that $\sigma_v(i)$ is the index of the $i^{th}$ highest degree neighbour of~$v$.

$S_v$ is then defined for $u \in \Gamma(v)$ as:
\begin{equation*}
S_v = \{u_{\sigma_v(1)}, ..., u_{\sigma_v(k_v)}\} 
\end{equation*}

We discuss the value to be given to $k_v$ in Section \ref{expes}.

The motivation behind this way of proceeding is that a vertex cannot know with certitude if it lies in the core of a dense area of the graph or in a more remote part, but can improve its knowledge with the help of information transmitted by its neighbours.

However, to avoid using massive propagation, which is the case for the label propagation based methods, we limit the information exchange, about the degree, to the immediate neighbourhood. Given the low value of diameter generally encountered in community graphs, especially inside the subgraph formed by a community, the immediate neighbourhood contains sufficient information to perform a fairly accurate detection.


\subsection{Agreement Computation}

Once the list has been established, the next step brings together
vertices that are most likely to form a community.  We propose to
compare the $S$ lists for each connected pair $(v, u)$,
ie $S_v$ and $S_u$ for $u \in \Gamma(v)$.

The idea is that if vertices share the same
neighbours in their $S_v$, they are more likely to be part of the same
community, because the are connected to the same high-degree vertices (common neighbourhood assumption, tied with the preferential attachment theory). We assume that a vertex
having fewer connection wants to join a vertex having more
connections.

We thus propose to compute the agreement (agree. in short) between two neighbour
vertices $u$ and $v$ as the number of vertices in common between $S_u$
and $S_v$:
\begin{equation*}
\textrm{agreement}(u,v) = |S_u \cap S_v|
\end{equation*}

We then propose to associate to each vertex $v$ its neighbour with
maximal agreement called $a_v$.

However, this choice is only relevant if the agreement is high enough, so
we use a parameter $\tau \in [0,1]$ to restrict the selection of $a_v$
among $v$'s neighbours having a $\tau\text{-percentage}$ of agreement
with it, imposing the condition that $u \in \Gamma(v), \textrm{agreement}(u,v) \geq
\tau\cdot \min(d_u,d_v)$. If none of $v$'s neighbours fulfil an agreement greater than $\tau$, meaning that we lack information about $v$ neighbourhood, we select the $v$'s neighbour of maximal degree as it is the best alternative.
So the expression of $a_v$ becomes:
\begin{equation*}
a_v = \begin{cases}
	\underset{u \in \Gamma(v)}{\argmax}\{\textrm{agree.}(u,v)~|~\textrm{agree.}(u,v) \geq \tau\cdot \min(d_u,d_v)\}
	\\
	\argmax\{d_u | u \in \Gamma(v)\}~\textrm{if the case above returns}~\varnothing
	\end{cases}
\end{equation*}

One of the main advantages is that, at the protocol level,
any vertex $v$ needs to broadcast
its list $S_v$ only to its neighbours for the agreement computation
to be made. This property is in line with the
desired limited propagation of the considered application framework.

This process is sketched in Algorithm~\ref{algo:agree}.

\begin{algorithm}[t]
\caption{Agreement computation and assignment}
\label{algo:agree}
\begin{algorithmic}[1]
	\REQUIRE vertex $v$, parameter $\tau \in [0,1]$
	\ENSURE vertex $a_v$, assigned vertex for $v$
	
	\STATE $v_{max} \gets \argmax\{$agreement$(u,v) | u \in \Gamma(v)\}$
	
	\IF{agreement$(v, v_{max}) \geq \tau\cdot \min(d_v,d_{v_{max}})$}
		\STATE $a_v \gets v_{max}$
	\ELSE
		\STATE $a_v \gets \argmin_{u \in \Gamma(v)} s(v,u)$
	\ENDIF	
\end{algorithmic}
\end{algorithm}
\begin{algorithm}[t]
\caption{Community uncovering}
\label{algo:uncover}
\begin{algorithmic}[1]
  \REQUIRE set of vertices $\mathcal{V}$, \ENSURE set of communities
  $\mathcal{C}$ 

  \STATE $\mathcal{C} = \{\{v\} | v \in \mathcal{V}\}$ \STATE
  \textbf{take} $v$ not processed yet \STATE ~~merge$(C(v), C(a_v))$
\end{algorithmic}
\end{algorithm}
\subsection{Community Uncovering}

The community detection finally exploits the extracted information,
namely the association of each vertex $v$ with its preferred vertex $a_v$.
The process is initialised to a
configuration where each vertex forms its own community. The next step
is to merge the community of each vertex $v$ with the one of its~$a_v$.

This process is sketched in Algorithm~\ref{algo:uncover}.

To do so and to minimise propagation we introduce an entity~$\epsilon$
able to poll easily a certain number of vertices from an area in the
graph, denoted $\mathcal{V} \subset V$. $\epsilon$ can be a high degree vertex or a
dedicated unit outside the network. For example, in a smartphone ad-hoc
network, $\epsilon$ can be a designated more powerful smartphone of the
network or a fixed computing unit if the network is located in a closed
space for example an exhibit.

$\epsilon$ then polls a vertex $v$ and merges its community with
the community of its associated vertex $a_v$.

The community uncovering process can be distributed on multiple
$\epsilon$ at the same time, thus the use of $\mathcal{V}$. As a
matter of fact, $\epsilon$ just needs to have access to every vertex
$v \in \mathcal{V}$, the merging being
deterministic. 

In order to process the whole graph, the only requirement is that the
union of all considered subsets $\mathcal{V}$ equals the whole set of vertices. No
condition regarding their intersection need to be applied, in
particular, they do not need to be disjoint.


\subsection{Algorithm Properties}
\subsubsection{Determinism}
Two out of three steps are deterministic: the lists of candidates only depend on the
degrees, so they remain the same for every instance of the
algorithm and any order of calculation. Merging is a symmetric
action, and the result does not depend on the order.

Agreement computation is not deterministic in itself; it is dependent
on the order in which the vertices are considered. Also, when several
candidates are suitable for an $a_v$, one is chosen at random. However our experimental tests
show that the variations and standard deviations are minimal.

\subsubsection{Complexity}
The list of candidates calculation process is run separately and
simultaneously by each vertex $v$ and depends on the sorting of its neighbours degree.
Therefore, the complexity of this process for a single vertex $v$ calculation with a $\mathcal{O}(n \cdot log~n)$ sorting algorithm can be evaluated as $\mathcal{O}(d_v log~d_v)$, and as
$\mathcal{O}(\bar{d}~log~\bar{d})$ on average for any $v \in V$, where $\bar{d}$
is the mean degree of the graph, for example $\frac{m}{n}$ for a
balanced random graph.

Assuming that the calculations to compile the lists of candidates are distributed over the
$n$ vertices, for the whole graph it takes $n \times \mathcal{O}(\bar{d}~log~\bar{d})$.

It is the same for the agreement computation, each vertex $v$ compares his $S_v$ list to those of its neighbours, yielding an $\mathcal{O}(d_v \cdot k_v)$ complexity, and $n \times \mathcal{O}(\bar{d} \cdot \bar{k})$ for the whole graph where $\bar{k}$ is the mean $k_v$ over all $v \in V$. Given that $\forall v \in V, k_v \leq d_v$, we can write the complexity as $n \times \mathcal{O}(\bar{d}^2)$. It is important to keep in mind that the calculations are run separately and simultaneously by all the vertices, thus we write the complexity that way.

Finally the community uncovering consists of a merge for each vertex considered by the process. Each process then runs in $\mathcal{O}(|\mathcal{V}|)$, and the complexity for the whole graph is $\mathcal{O}(n)$.

\subsubsection{Overlapping communities}
\label{overlap}In the first version of this method, we choose not to consider overlapping communities, because evaluating the quality of community structures with multiple belonging vertices is harder than those with single belonging vertices.

However, a simple way to enable the detection of overlapping communities would be to make $a_v$ a set and define the agreement procedure accordingly. So a vertex $v$ would join the community of each vertex in the $a_v$ set, instead of joining the community of a single preferred vertex $a_v$.

\section{Experiments}
\label{expes}
This section discusses the results of experiments both on artificial
and real-world graphs, to show the relevance of the proposed method,
after presenting the considered quality
criteria and parameter values.

\subsection{Quality Criteria}
We evaluate the obtained results using ARI and NMI, criteria
usually applied to assess the results of clustering tasks, to compare
partitions of a given set of objects.


The Adjusted Rand Index (ARI) is an enhancement of the
original Rand Index corrected for chance
\cite{hubert_comparing_1985}. It produces a value between $-1$ (fully
incorrect match between two partitions) and $1$ (fully exact match
between partitions), $0$ being, on average, the value for the
comparison of two partitions obtained at random.
	
The Normalized Mutual Information (NMI) reflects the amount
of information contained in the considered partition compared to the
ground truth, normalised by the average of Shannon's entropy of the two
partitions~\cite{strehl_cluster_2003}:
$0$ is the value obtained for a random assignment and $1$ for a fully
similar one. We use here the word ``similar'' because there can be a
higher number of clusters in the produced result than in the ground
truth, but NMI remains high as long as the obtained partition remains
close to the ground truth, even if the classifier produces several
clusters instead of one.
	 

We consider ARI as a measure of whether the found communities are
exactly the same (same number, same vertex partition) thus
decreasing when the number of clusters differs, and NMI as a measure
of how the clusters are meaningful with respect to the ground truth,
ie that globally the vertices tend to be put together in both
partitions even if the number of clusters is different.

\subsection{Parameter Selection}

The proposed algorithm depends on two parameters, $k_v$ and $\tau$. 

We set $k_v = \max(1, \frac{d_v}{2})$, as an intermediary value. Obviously,
setting $k=1$ is insufficient: a vertex would choose its $a_v$ to be its highest degree neighbour, but if this
highest degree neighbour is not in its community, for example because $v$ lies in the
border area of its community, intuitively as much linked to other communities as to
vertices of its own community, it will not be placed in the most appropriate community. We experimentally found that it happens frequently, distorting  the vertex partition.

On the other hand, keeping all the
neighbours is not necessary: communities being considered here as
denser subgraphs, a vertex is connected to the majority of its
neighbours with high probability (high clustering coefficient).
Moreover, the most significant neighbours bringing information,
those of highest degrees, represent a small part of a vertex neighbourhood.
As a matter of fact, the degrees in the whole graph are expected to follow a power-law
distribution as it is really common in community networks especially social networks \cite{barabasi_emergence_1999}.

We experimentally set $\tau = 0.2$: it appears to be the first value
that achieves the best NMI, which is not improved for lower
values of $\tau$.

\subsection{Artificial Graphs}
Artificially generated graphs allow a ground truth comparison as their
properties are known by construction. We consider two generation
procedures, one similar to Clauset's \cite{clauset_finding_2005} and
the well-known LFR benchmark \cite{lancichinetti_benchmark_2008}.

\subsubsection{Comparison with a local approach}
We start here with a comparison with another local community identification
method, from Clauset. We run the same kind of
experiment that he did in his paper \cite{clauset_finding_2005}.

Clauset uses generated graphs of $n = 128$ vertices
divided into four communities of the same size, 32 vertices each, which
constitutes the ground truth. Each
vertex expected degree, $z$, is the sum of the intra-community edges
$z_{in}$ and inter-community edges $z_{out}$. The edges are placed
independently at random between vertices so that $z, z_{in}$ and
$z_{out}$ are, in expectation, respected. Clauset sets $z = 16$ and
makes $z_{out}$ vary, so 
$z_{in} = z - z_{out}$.

We compare the performance of our method to that of Clauset's local modularity on the whole graph.
Indeed, Clauset's algorithm starts from a seed vertex in the graph and returns a community around that seed vertex, so it does not always process the whole graph at once. Clauset runs four detections processes, one for each community, and then compares the results community by community against the ground truth.

The algorithm we propose being applied to the whole graph, we had to find a way to aggregate Clauset's algorithm results for the whole graph as well. First, we choose to run Clauset's algorithm for each community of each graph, that is four times for each graph, selecting the seed vertex as the one having the highest degree in each community. That's a main advantage given to Clauset's method because it facilitates the detection, telling it where to start and omitting the fact that this method, contrary to the one we propose, cannot detect multiple communities at the same time in the graph.

Then, we aggregate the results as follows: a vertex classified in several communities is kept in only one community selected at random (as we do not consider overlapping communities here), the unclassified vertices are each put in a self-community in which they are the sole vertex, allowing a fair comparison with our method, where every vertex is assigned to only one community.

\begin{figure}[t]
	\centering
	\includegraphics[width=0.45\textwidth]{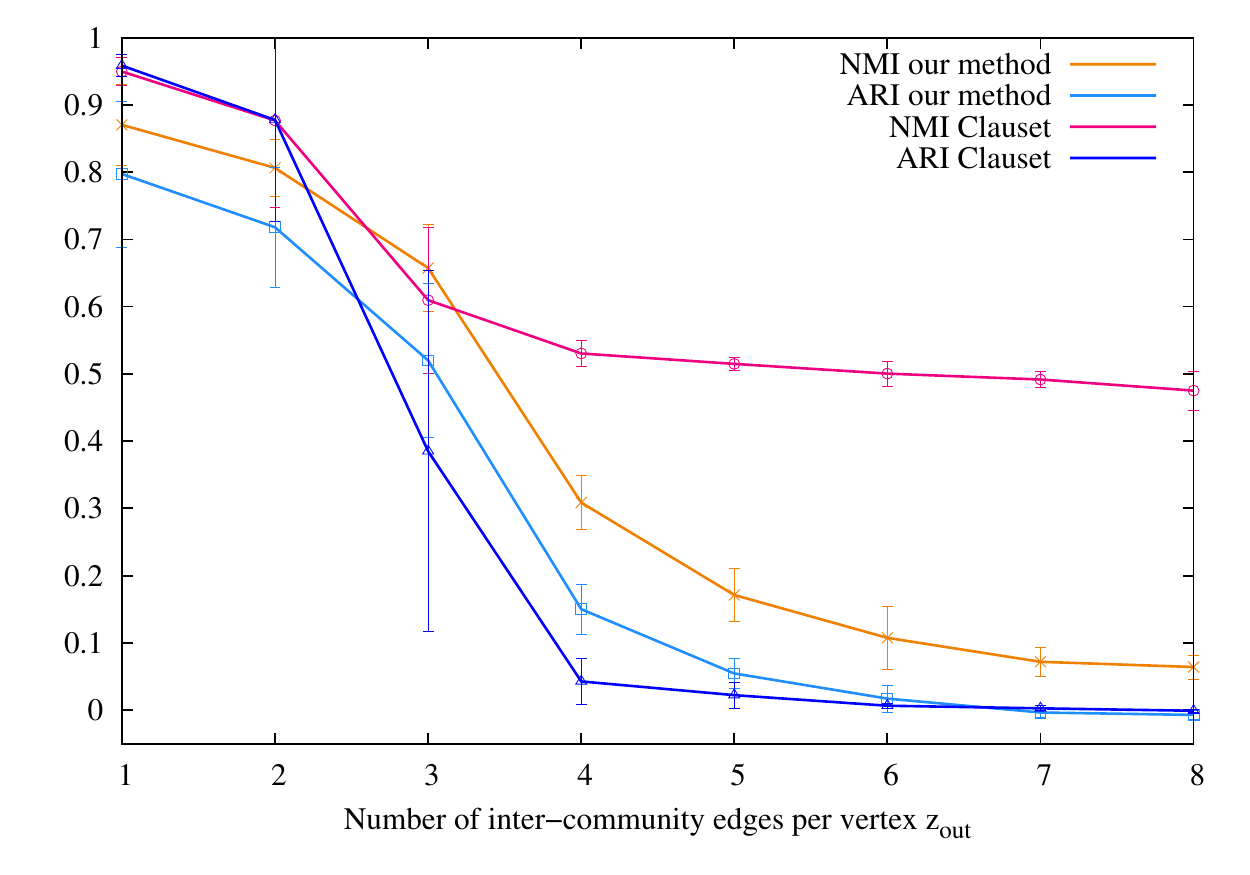}
	\caption{Average ARI and NMI for all communities over 500
          generations, as compared to ground truth, for Clauset's like
          graphs \cite{clauset_finding_2005}, as a function of
          inter-community edges $z_{out}$. }
	\label{fig:clauset_nmi}
\end{figure}

The experiment is performed for $z_{out}$ from 1 to 8, over 10 different graphs for each value of $z_{out}$. The results are presented in Figure~\ref{fig:clauset_nmi}. It can be observed that the performance of the method we propose is not as good as Clauset's, except for $z_{out} = 3$.

However even if the NMI remains higher for Clauset's method than the one we propose, the ARI shows that the community partition found by Clauset is not very similar to the original one. That is because Clauset's method classifies many vertices in several communities or not at all, so the agregation applied smoothes these results by removing the multi-affectations and adding the missing vertices, producing many single-vertex communities that do not affect much the NMI, contrary to the ARI. Our interpretation is that Clauset's method detects well the core of each community but gets quickly lost when it goes toward the boundaries.

On the contrary, the performance of the method we propose gradually decreases but the ARI remains proportional to the NMI, showing that the meaningfulness of the community partition, compared with the original, is gradually lost and tends to resemble randomness when $z_{out} \geq 7$. As a matter of fact, the ARI compared to the original partition is more favourable for the method we propose when $z_{out} = $ 4 and 5 whereas the NMI is more favourable for Clauset's method.

Also, we can notice that the standard deviation for Clauset's method is high when $z_{out} = $ 2 and 3, before the performance drop, whereas the standard deviation for ours remains moderate. We interpret this as a notable instability of Clauset's method when there is significant noise, i.e. when $z_{out} \geq 2$, leading to the performance drop.


\subsubsection{Comparison with global approaches}
The Lancichinetti,Fortunato,Radicchi (LFR) random graphs generator \cite{lancichinetti_benchmark_2008} allows
to produce better community graphs with regards to the properties of
such graphs \cite{girvan_community_2002}, than models similar to the one
used by Clauset. In particular, the parameter $\mu \in [0,1]$ indicates
how "well-knit" the graph is, i.e. how much the
community are clearly separated, therefore easily identifiable.

First, we run an experiment to set the best value to use for the parameter $\tau$,
then we compare the method we propose to two state-of-the-art global methods:
Louvain~\cite{blondel_fast_2008} and InfoMap~\cite{rosvall_maps_2008}.

We consider a graph with 1000 vertices for each value of~$\mu$ from
$0.1$ to $0.8$ increasing by $0.1$, generated by LFR, and test the algorithm
we propose for different values of $\tau$.

\begin{figure*}	
       \subfloat[NMI as a function of mixing parameter $\mu$]{
       		\includegraphics[width=0.45\textwidth]{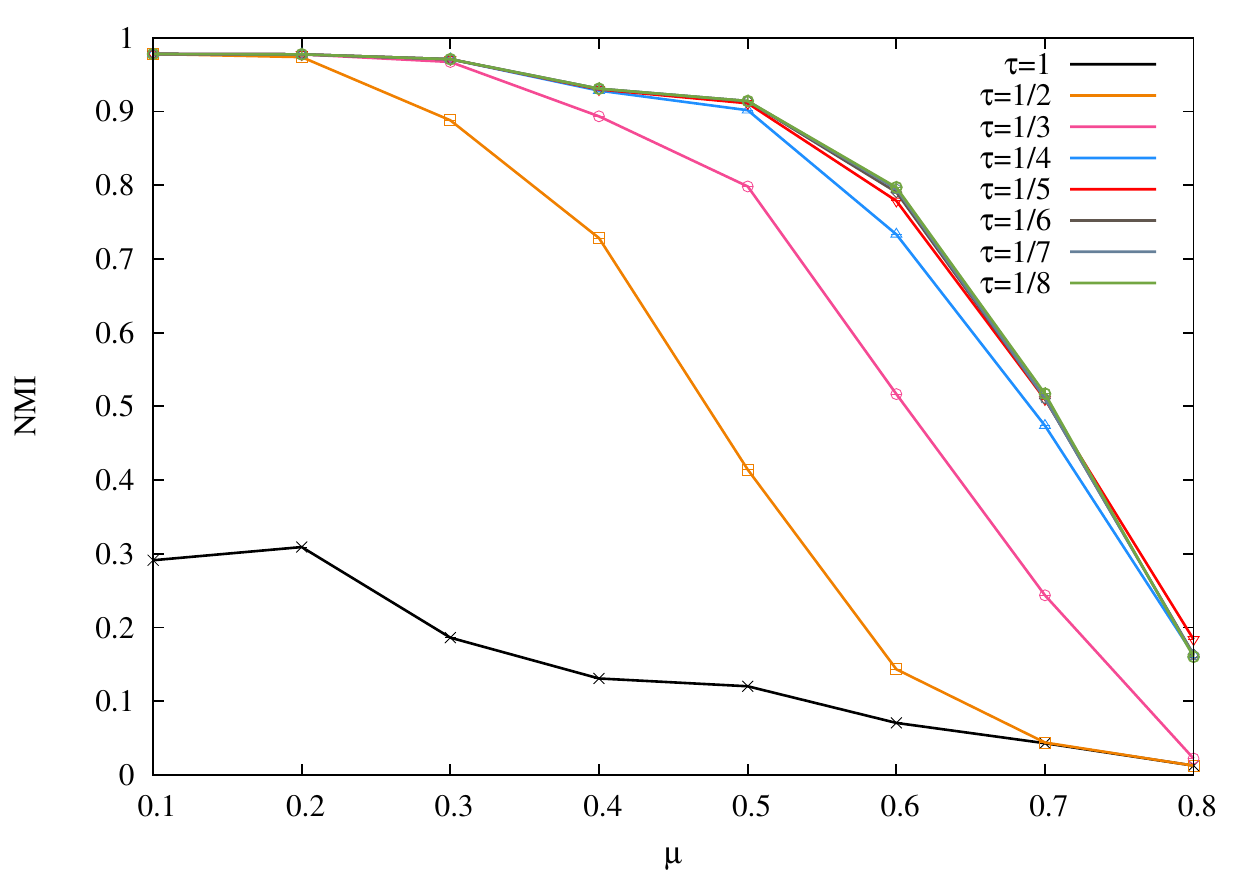}
		}
		\vspace{0.15in}
		\subfloat[ARI as a function of mixing parameter $\mu$]{
       		\includegraphics[width=0.45\textwidth]{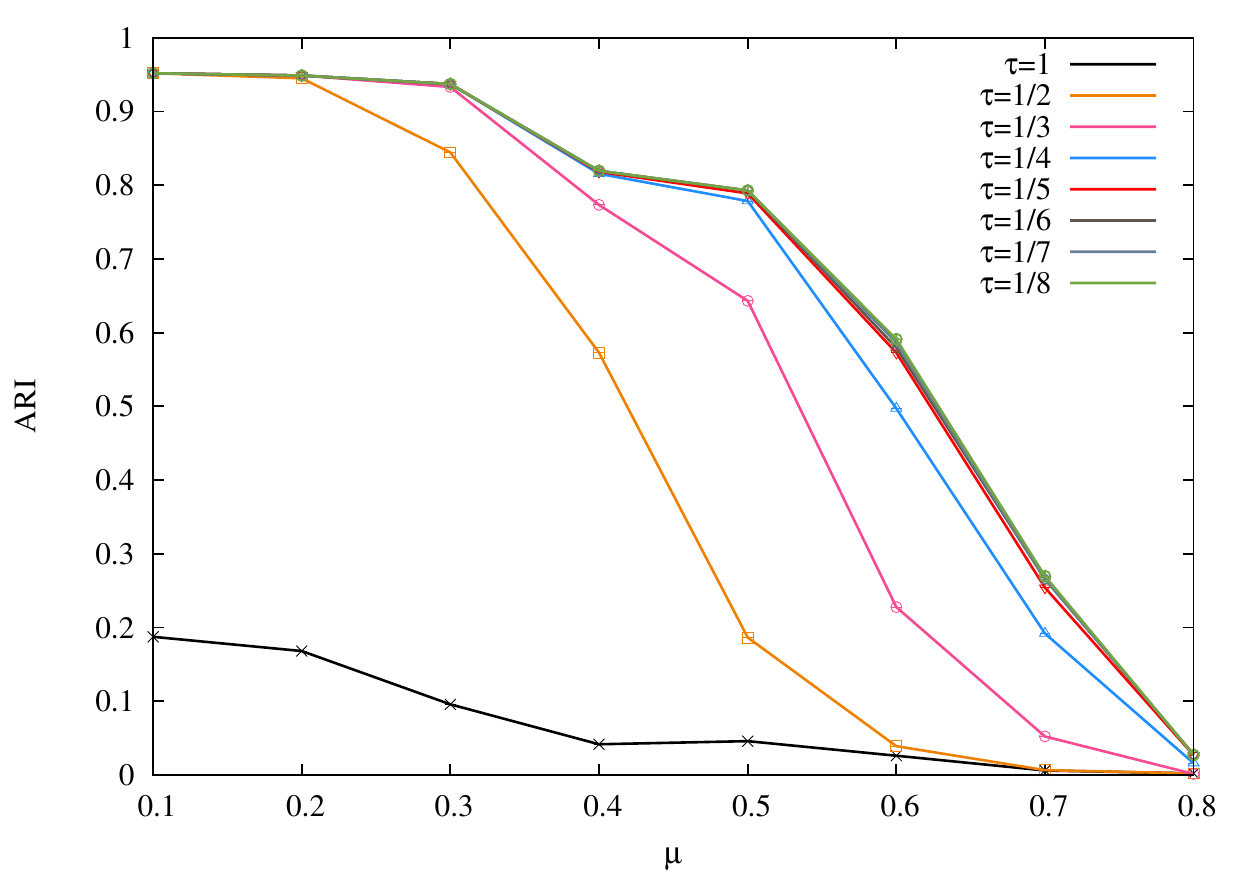}
		}
		\caption[caption]{Evaluation of our method on LFR-benchmark generated graphs (1000 vertices), for different values of $\tau$.\\\hspace{\textwidth}
		Curves for $\tau = \frac{1}{6}$ to $\frac{1}{8}$ are too close from each other and appear combined}
		\label{fig:lfr_taus}
\end{figure*}

Figure~\ref{fig:lfr_taus} shows the obtained the ARI and the NMI, mean and
standard deviation averaged over 100 runs.

We can see that the method we propose performs well until $\mu = 0.2$, decreasing
for higher values of $\mu$, a sign of instability due to the
fact that the communities become less and less identifiable.

The NMI stays higher than the ARI, meaning that for a given $\mu$, although
the method we propose returns a different number of clusters than the
ground truth, the partition of the vertices remains closer to the
ground truth partition.

\begin{figure*}	
       \subfloat[NMI as a function of mixing parameter $\mu$]{
       		\includegraphics[width=0.45\textwidth]{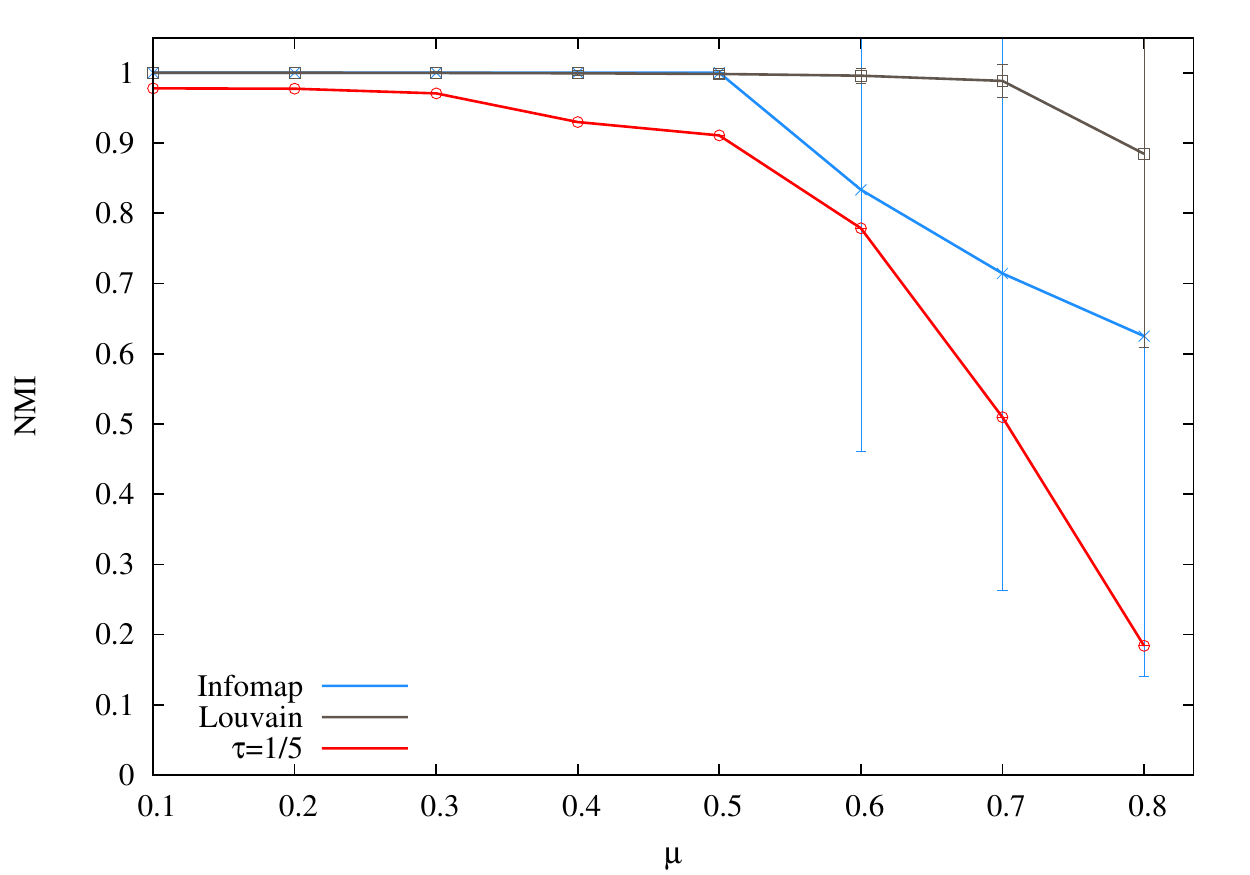}
		}
		\vspace{0.15in}
		\subfloat[ARI as a function of mixing parameter $\mu$]{
       		\includegraphics[width=0.45\textwidth]{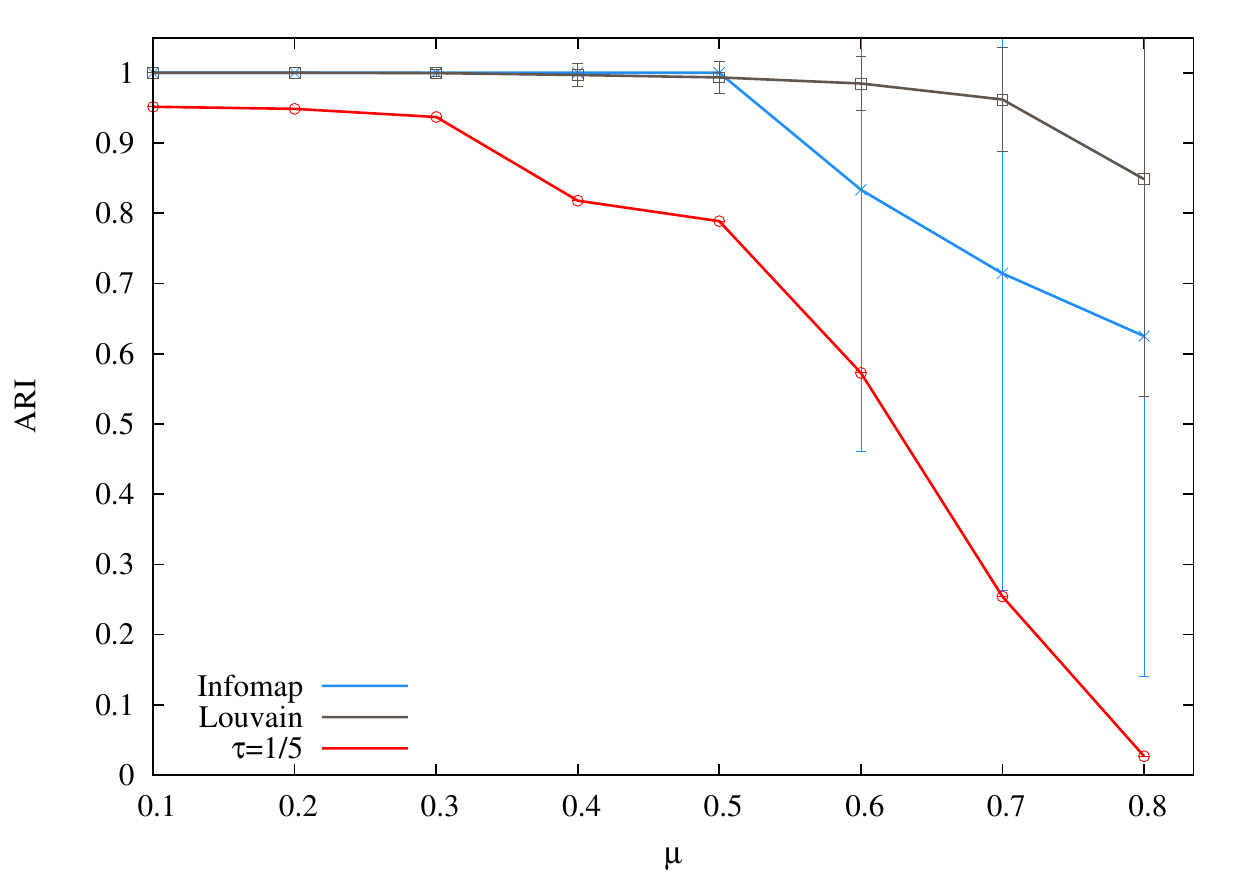}
		}
		\caption{Evaluation of our method on LFR-benchmark generated graphs (1000 vertices), for different methods}
		\label{fig:lfr_vs}
\end{figure*}

Then, we compare the method we propose to Louvain and InfoMap
using the value $\tau = 0.2$ that yields the best results. The
results, presented on Figure~\ref{fig:lfr_vs}, show that the NMI stays
over 0.9 until $\mu = 0.5$. Louvain performs better here compared to InfoMap,
which is not always the case \cite{hric_community_2014}, but both see
their standard deviation increase, especially InfoMap for $\mu >
0.5$. Although the quality of the results produced by the method we propose
decreases, the advantage here is its stability (in addition to its
distributed mode of operation, cf. Section \ref{algo}).

Globally, as long as the community structure remains sufficiently
defined ($\mu \leq 0.5$), our algorithm performs quite well as
compared to global methods.

\subsection{Real-world Graphs}
Contrary to artificial graphs, real-world graphs are not generated after a given model but are the encoding of a real situation. They come with no ground truth, even if some of there properties are known, so the ground truth have to be human-labelled or generated by a community detection method.


\subsubsection{Zachary's Karate Club}
Zachary's karate club is a real situation
\cite{zachary_information_1977} discussing over the conflicts leading
to the separation of a karate club into two factions. The interactions between members
were observed for 3 years from a sociological/ethnographic point of
view. After the splitting of the initial club, the new partition was found to be strongly correlated with
the interactions over the past preceding years (that was the purpose
of Zachary's paper). The two factions are seen as the 2 main communities
which the graph can be divided into. It is a well-known small
benchmark to study the behaviour of a community detection algorithm.

\begin{figure}[t]
	\centering
	\includegraphics[width=0.4\textwidth]{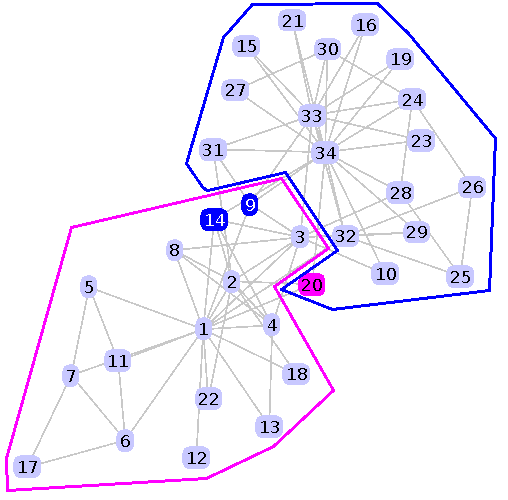}
	\caption{Community partition for Zachary's karate club. The highlighted vertices are the only ones not compliant with the original partition.}
	\label{fig:zachary}
\end{figure}

Figure \ref{fig:zachary} shows the partition made by our
algorithm. The NMI and ARI for this case are 0.65 and 0.67
respectively. Three vertices are not put in the expected community
compared to the partition stated by Zachary : \#9, \#14 and \#20
(highlighted on the figure). These vertices are at the interface
between the two communities and are all connected to vertex \#34, one
of the most influent in this community (highest degree). Therefore the
classification does not seem absurd.

%
%
%

\subsubsection{Amazon}
We also apply the method we propose on a co-purchasing network from Amazon.com, collected by Yang and Leskovec \cite{yang_defining_2013} and available online \cite{leskovec_snap_2014}. The graph has $n = 334,863$ vertices and $m = 925,872$ edges. It is undirected and has no isolated vertex nor self-loop. It has a clustering coefficient of approximately $0.40$, which is really low (neighbours are sparsely connected, there are few triangles).
Each vertex represents a product sold on Amazon.com and each edge between two products means that they are frequently co-purchased.

The main advantage of this dataset is that it comes with a ground truth, i.e. it is possible and relevant to apply the method we propose and compare the result with the ground truth communities.

We removed from the results of our detection the vertices having no affected ground truth community, and we kept only one community at random for vertices belonging to several communities.

We achieve a NMI of 0.905, in other words our method is not only able
to find communities in computer-generated graphs with controlled
properties, but also in large graphs modeling -or taken from- real
situations.

\section{Conclusion}
\label{conclusion}
We presented a decentralised method to detect communities on a graph,
using lists based on the degrees of the vertices. The
algorithm, divided in 3 steps, makes each vertex elaborate a list of defined
vertices, then an agreement with the list of its neighbours. A process
is run to put together vertices that have the highest agreement.

This method relies on simple calculations with few inputs, and is
therefore well adapted to detect communities during a live event,
where each participant carries a wireless mobile device
(e.g. smartphone, smartwatch) that can do the computation and share
information with other nearby devices.

We showed that this method performs well both on artificial and real
graphs, although less than state-of-the-art algorithms. However,
a main advantage is that it is designed to be run on a distributed
architecture, more precisely over a network of connected ``smart" devices e.g.
smartphones. Furthermore, most of the method is deterministic, giving rather stable results, more easily predictable and more explainable.

We plan to do further experimentations with the proposed method on a larger
number of graphs of various properties and models, to better
understand the limits of the assignment based on agreement. Future
works will also include experiments on dynamic graphs, as we think
that this method is easily adaptable to the dynamic case. 

Perspective also include an extension of the proposed method to the
case of overlapping communities (cf. \ref{overlap}) and attributed graphs:
adding other information available to constitute the lists, such as the profile
of a participant carrying the device in the considered context, would greatly
improve the quality of the community formation.

\section*{Acknowledgments}
This work was performed as part of the Homo Textilus project, supported by the French National Agency for Research (ANR) under the grant ANR-11-SOIN-007.
%
\IEEEpeerreviewmaketitle


\bibliographystyle{IEEEtran}
\bibliography{IEEEabrv,bib_asonam}

\begin{thebibliography}{10}
\providecommand{\url}[1]{#1}
\csname url@samestyle\endcsname
\providecommand{\newblock}{\relax}
\providecommand{\bibinfo}[2]{#2}
\providecommand{\BIBentrySTDinterwordspacing}{\spaceskip=0pt\relax}
\providecommand{\BIBentryALTinterwordstretchfactor}{4}
\providecommand{\BIBentryALTinterwordspacing}{\spaceskip=\fontdimen2\font plus
\BIBentryALTinterwordstretchfactor\fontdimen3\font minus
  \fontdimen4\font\relax}
\providecommand{\BIBforeignlanguage}[2]{{%
\expandafter\ifx\csname l@#1\endcsname\relax
\typeout{** WARNING: IEEEtran.bst: No hyphenation pattern has been}%
\typeout{** loaded for the language `#1'. Using the pattern for}%
\typeout{** the default language instead.}%
\else
\language=\csname l@#1\endcsname
\fi
#2}}
\providecommand{\BIBdecl}{\relax}
\BIBdecl

\bibitem{wasserman_social_1994}
S.~Wasserman and K.~Faust, \emph{\BIBforeignlanguage{English}{Social {Network}
  {Analysis}: {Methods} and {Applications}}}, 1st~ed.\hskip 1em plus 0.5em
  minus 0.4em\relax Cambridge: Cambridge University Press, Nov. 1994.

\bibitem{chaintreau_impact_2007}
A.~Chaintreau, P.~Hui, J.~Crowcroft, C.~Diot, R.~Gass, and J.~Scott, ``Impact
  of {Human} {Mobility} on {Opportunistic} {Forwarding} {Algorithms},''
  \emph{IEEE Transactions on Mobile Computing}, vol.~6, no.~6, pp. 606--620,
  Jun. 2007.

\bibitem{vastardis_mobile_2013}
N.~Vastardis and K.~Yang, ``Mobile {Social} {Networks}: {Architectures},
  {Social} {Properties}, and {Key} {Research} {Challenges},'' \emph{IEEE
  Communications Surveys Tutorials}, vol.~15, no.~3, pp. 1355--1371, 2013.

\bibitem{lee_mobility_2013}
K.~Lee, P.~Hui, and S.~Chong, ``\BIBforeignlanguage{en}{Mobility {Models} in
  {Opportunistic} {Networks}},'' in \emph{\BIBforeignlanguage{en}{Mobile {Ad}
  {Hoc} {Networking}}}, S.~Basagni, M.~Conti, S.~Giordano, and I.~Stojmenovic,
  Eds.\hskip 1em plus 0.5em minus 0.4em\relax John Wiley \& Sons, Inc., 2013,
  pp. 360--418.

\bibitem{hui_distributed_2007}
P.~Hui, E.~Yoneki, S.~Y. Chan, and J.~Crowcroft, ``Distributed {Community}
  {Detection} in {Delay} {Tolerant} {Networks},'' in \emph{Proc. of the 2007
  {ACM}/{IEEE} {International} {Workshop} on {Mobility} in the {Evolving}
  {Internet} {Architecture}}, ser. {MobiArch} '07.\hskip 1em plus 0.5em minus
  0.4em\relax New York, NY, USA: ACM, 2007, pp. 7:1--7:8.

\bibitem{hui_bubble_2011}
P.~Hui, J.~Crowcroft, and E.~Yoneki, ``{BUBBLE} {Rap}: {Social}-{Based}
  {Forwarding} in {Delay}-{Tolerant} {Networks},'' \emph{IEEE Transactions on
  Mobile Computing}, vol.~10, no.~11, pp. 1576--1589, Nov. 2011.

\bibitem{laibowitz_sensor_2006}
M.~Laibowitz, J.~Gips, R.~Aylward, A.~Pentland, and J.~Paradiso, ``A sensor
  network for social dynamics,'' in \emph{Proc. of the 2006 {Information}
  {Processing} in {Sensor} {Networks} ({ISPN})}, 2006, pp. 483--491.

\bibitem{paradiso_identifying_2010}
J.~A. Paradiso, J.~Gips, M.~Laibowitz, S.~Sadi, D.~Merrill, R.~Aylward,
  P.~Maes, and A.~Pentland, ``Identifying and {Facilitating} {Social}
  {Interaction} with a {Wearable} {Wireless} {Sensor} {Network},''
  \emph{Personal Ubiquitous Computing}, vol.~14, no.~2, pp. 137--152, 2010.

\bibitem{radicchi_defining_2004}
F.~Radicchi, C.~Castellano, F.~Cecconi, V.~Loreto, and D.~Parisi,
  ``\BIBforeignlanguage{en}{Defining and identifying communities in
  networks},'' \emph{\BIBforeignlanguage{en}{Proc. of the National Academy of
  Sciences}}, vol. 101, no.~9, pp. 2658--2663, Mar. 2004.

\bibitem{fortunato_community_2009}
S.~Fortunato, ``Community detection in graphs,'' \emph{Physics Reports-Review
  Section of Physics Letters}, pp. 75--174, Jun. 2009.

\bibitem{kernighan_efficient_1970}
B.~W. Kernighan and S.~Lin, ``\BIBforeignlanguage{en}{An {Efficient}
  {Heuristic} {Procedure} for {Partitioning} {Graphs}},''
  \emph{\BIBforeignlanguage{en}{Bell System Technical Journal}}, vol.~49,
  no.~2, pp. 291--307, 1970.

\bibitem{bron_algorithm_1973}
C.~Bron and J.~Kerbosch, ``Algorithm 457: finding all cliques of an undirected
  graph,'' \emph{Communications of the ACM}, vol.~16, no.~9, pp. 575--577, Sep.
  1973.

\bibitem{freeman_set_1977}
L.~Freeman, ``A {Set} of {Measures} of {Centrality} {Based} on {Betweenness},''
  \emph{Sociometry}, vol.~40, no.~1, pp. 35--41, Mar. 1977.

\bibitem{girvan_community_2002}
M.~Girvan and M.~E.~J. Newman, ``\BIBforeignlanguage{en}{Community structure in
  social and biological networks},'' \emph{\BIBforeignlanguage{en}{Proc. of the
  National Academy of Sciences}}, vol.~99, no.~12, pp. 7821--7826, Jun. 2002.

\bibitem{newman_finding_2004}
M.~E.~J. Newman and M.~Girvan, ``Finding and evaluating community structure in
  networks,'' \emph{Physical Review E}, vol.~69, no.~2, p. 026113, 2004.

\bibitem{clauset_finding_2004}
A.~Clauset, M.~E.~J. Newman, and C.~Moore, ``Finding community structure in
  very large networks,'' \emph{Physical Review E}, vol.~70, no.~6, p. 066111,
  2004.

\bibitem{blondel_fast_2008}
V.~D. Blondel, J.-L. Guillaume, R.~Lambiotte, and E.~Lefebvre,
  ``\BIBforeignlanguage{en}{Fast unfolding of communities in large networks},''
  \emph{\BIBforeignlanguage{en}{Journal of Statistical Mechanics: Theory and
  Experiment}}, vol. 2008, no.~10, p. P10008, Oct. 2008.

\bibitem{aldecoa_deciphering_2011}
R.~Aldecoa and I.~Mar\'in, ``Deciphering {Network} {Community} {Structure} by
  {Surprise},'' \emph{PLoS ONE}, vol.~6, no.~9, p. e24195, Sep. 2011.

\bibitem{fortunato_resolution_2007}
S.~Fortunato and M.~Barth\'elemy, ``Resolution limit in community detection,''
  \emph{Proc. of the National Academy of Sciences}, vol. 104, no.~1, pp.
  36--41, 2007.

\bibitem{hoff_latent_2002}
P.~D. Hoff, A.~E. Raftery, and M.~S. Handcock, ``Latent {Space} {Approaches} to
  {Social} {Network} {Analysis},'' \emph{Journal of the American Statistical
  Association}, vol.~97, no. 460, pp. 1090--1098, 2002.

\bibitem{rosvall_maps_2008}
M.~Rosvall and C.~T. Bergstrom, ``\BIBforeignlanguage{en}{Maps of random walks
  on complex networks reveal community structure},''
  \emph{\BIBforeignlanguage{en}{Proc. of the National Academy of Sciences}},
  vol. 105, no.~4, pp. 1118--1123, Jan. 2008.

\bibitem{cazabet_simulate_2011}
R.~Cazabet and F.~Amblard, ``Simulate to {Detect}: {A} {Multi}-agent {System}
  for {Community} {Detection},'' in \emph{Proc. of the 2011 {IEEE}/{WIC}/{ACM}
  {Web} {Intelligence} and {Intelligent} {Agent} {Technology} ({WI}-{IAT})},
  vol.~2, 2011, pp. 402--408.

\bibitem{hui_pocket_2005}
P.~Hui, A.~Chaintreau, J.~Scott, R.~Gass, J.~Crowcroft, and C.~Diot, ``Pocket
  {Switched} {Networks} and {Human} {Mobility} in {Conference}
  {Environments},'' in \emph{Proc. of the 2005 {ACM} {SIGCOMM} {Workshop} on
  {Delay}-tolerant {Networking}}, ser. {WDTN} '05.\hskip 1em plus 0.5em minus
  0.4em\relax New York, NY, USA: ACM, 2005, pp. 244--251.

\bibitem{clauset_finding_2005}
A.~Clauset, ``Finding local community structure in networks,'' \emph{Physical
  Review E}, vol.~72, no.~2, p. 026132, Aug. 2005.

\bibitem{raghavan_near_2007}
U.~N. Raghavan, R.~Albert, and S.~Kumara, ``Near linear time algorithm to
  detect community structures in large-scale networks,'' \emph{Physical Review
  E}, vol.~76, no.~3, p. 036106, Sep. 2007.

\bibitem{kanawati_seed-centric_2014}
R.~Kanawati, ``Seed-centric approaches for community detection in complex
  networks,'' in \emph{Social {Computing} and {Social} {Media}}.\hskip 1em plus
  0.5em minus 0.4em\relax Springer, 2014, pp. 197--208.

\bibitem{barabasi_emergence_1999}
A.-L. Barab\'asi and R.~Albert, ``\BIBforeignlanguage{en}{Emergence of
  {Scaling} in {Random} {Networks}},'' \emph{\BIBforeignlanguage{en}{Science}},
  vol. 286, no. 5439, pp. 509--512, Oct. 1999.

\bibitem{hubert_comparing_1985}
L.~Hubert and P.~Arabie, ``\BIBforeignlanguage{en}{Comparing partitions},''
  \emph{\BIBforeignlanguage{en}{Journal of Classification}}, vol.~2, no.~1, pp.
  193--218, Dec. 1985.

\bibitem{strehl_cluster_2003}
A.~Strehl and J.~Ghosh, ``Cluster {Ensembles} — a {Knowledge} {Reuse}
  {Framework} for {Combining} {Multiple} {Partitions},'' \emph{Journal of
  Machine Learning Research}, vol.~3, pp. 583--617, Mar. 2003.

\bibitem{lancichinetti_benchmark_2008}
A.~Lancichinetti, S.~Fortunato, and F.~Radicchi, ``Benchmark graphs for testing
  community detection algorithms,'' \emph{Physical Review E}, vol.~78, no.~4,
  p. 046110, Oct. 2008.

\bibitem{hric_community_2014}
D.~Hric, R.~K. Darst, and S.~Fortunato, ``Community detection in networks:
  {Structural} communities versus ground truth,'' \emph{Physical Review E},
  vol.~90, no.~6, p. 062805, Dec. 2014.

\bibitem{zachary_information_1977}
W.~Zachary, ``An information flow model for conflict and fission in small
  groups,'' \emph{Journal of Anthropological Research}, vol.~33, no.~4, pp.
  452--473, 1977.

\bibitem{yang_defining_2013}
J.~Yang and J.~Leskovec, ``\BIBforeignlanguage{en}{Defining and evaluating
  network communities based on ground-truth},''
  \emph{\BIBforeignlanguage{en}{Knowledge and Information Systems}}, vol.~42,
  no.~1, pp. 181--213, Oct. 2013.

\bibitem{leskovec_snap_2014}
\BIBentryALTinterwordspacing
J.~Leskovec and A.~Krevl, ``{SNAP} {Datasets}: {Stanford} {Large} {Network}
  {Dataset} {Collection},'' Jun. 2014. [Online]. Available:
  \url{http://snap.stanford.edu/data}
\BIBentrySTDinterwordspacing

\end{thebibliography}

\end{document}